\documentclass[reprint, aip,jcp, superscriptaddress]{revtex4-2}

\usepackage{amsfonts}
\usepackage{amsmath}
\usepackage{amssymb}
\usepackage{graphicx}
\usepackage{dcolumn}
\usepackage{bm}
\usepackage{float}
\usepackage{amsthm}
\usepackage[urlcolor=white]{hyperref}
\usepackage{xcolor}

\graphicspath{{./Figures/}{"C:/Users/GROUP
LCPT/Documents/Fabian/HK_coherence_thermofield_dynamics/Figures/"}}

\begin{document}

\title{Vibronic spectra at nonzero temperatures from Herman--Kluk coherence thermofield dynamics}
\author{Fabian Kr\"{o}ninger}
\affiliation{Laboratory of Theoretical Physical Chemistry, Institut des Sciences et
Ing\'enierie Chimiques, Ecole Polytechnique F\'ed\'erale de Lausanne (EPFL),
CH-1015 Lausanne, Switzerland}
\author{Caroline Lasser}
\affiliation{Department of Mathematics, School of Computation,
Information and Technology, Technische Universit\"at M\"unchen, Germany.}
\author{Ji\v{r}\'{\i} J.~L. Van\'{\i}\v{c}ek}
\email{jiri.vanicek@epfl.ch} 
\affiliation{Laboratory of Theoretical Physical Chemistry, Institut des Sciences et
Ing\'enierie Chimiques, Ecole Polytechnique F\'ed\'erale de Lausanne (EPFL),
CH-1015 Lausanne, Switzerland}

\date{\today}
\begin{abstract}
We combine the semiclassical Herman--Kluk approximation with the coherence thermofield dynamics in order to evaluate vibrationally resolved electronic spectra at nonzero temperatures.
In coherence thermofield dynamics, the dipole time correlation function is rewritten exactly as a wavepacket autocorrelation function, and the corresponding wavepacket is a solution to a zero-temperature time-dependent Schr\"odinger equation on an augmented configuration space of doubled dimension.
We derive the Herman--Kluk representation for the thermofield wavepacket autocorrelation function and demonstrate how it can be computed from individual trajectories.
To analyze this method, we compare spectra of Morse potentials of increasing anharmonicity evaluated at various temperatures with a numerically exact approach, with the Herman--Kluk coherence thermofield dynamics, and with the single-trajectory thawed Gaussian coherence thermofield dynamics.
At low anharmonicity, both approximate methods yield accurate spectra.
However, in a Morse potential with higher anharmonicity, the thawed Gaussian thermofield dynamics, based on the local harmonic approximation, fails to capture emerging hot bands, whereas the  Herman--Kluk thermofield approach successfully reproduces them.
\end{abstract}

\maketitle
 \newpage

\section{Introduction}
In the time-dependent approach to spectroscopy, \cite{Heller:1981a, book_Mukamel:1999, book_Tannor:2007, book_Heller:2018} the computation of spectra is formulated as a molecular quantum dynamics problem.
Specifically, within time-dependent perturbation theory, the vibrationally resolved electronic spectrum is obtained from the Fourier transform of a dipole time correlation function.

Within the zero-temperature and Condon approximations, the dipole time correlation function reduces to a wavepacket autocorrelation function and exact quantum methods \cite{Feit_Steiger:1982, Meyer_Cederbaum:1990, Ben-Nun_Martinez:1999, Worth_Burghardt:2004} can be used for propagating the pure state.
Because the main features of a vibrationally resolved electronic spectrum are primarily determined by the short-time behavior of the nuclear wavepacket, approximate but efficient semiclassical methods \cite{Vleck:1928, Heller:1975, Heller:1976, Heller:1981, Herman_Kluk:1984, Coalson_Karplus:1990, book_Heller:2018, Begusic_Vanicek:2019, Vanicek:2023} are well-suited for numerical simulations.
Single-trajectory semiclassical approaches, such as the thawed Gaussian approximation, \cite{Heller:1975, Vanicek:2023} 
were developed to model weakly anharmonic systems with sufficient accuracy and low computational cost.
In systems with significant anharmonicity, where single-trajectory methods are inadequate, the multi-trajectory Herman--Kluk propagator \cite{Herman_Kluk:1984} often yields accurate spectra and provides physically meaningful insights because of its high accuracy. \cite{Swart_Rousse:2008, Robert:2010, Lasser_Lubich:2020}

Nonetheless, because experiments are often conducted at room or higher temperatures, modeling experimental results accurately requires accounting for thermal effects. 
The Herman--Kluk approximation has been used successfully for nonzero-temperature simulations of molecular infrared spectra, \cite{Sun_Miller:1998, Wang_Miller:1998, Sun_Miller:1999, Gelabert_Miller:2001, Miller:2001, Wang_Miller:2001a, Sun_Miller:2002, Zhao_Miller:2002, Tao_Miller:2011, Makri:2002, Makri_Miller:2002, Kegerreis_Makri:2007, Church_Ananth:2019, Lanzi_Conte:2024, Lanzi_Conte:2025} 
but its application to vibrationally resolved electronic spectra simulations at nonzero temperatures has received limited attention. \cite{Walton_Manolopoulos:1995, Tatchen_Pollak:2009}
Standard approaches for evaluating vibronic spectra at nonzero temperatures include Boltzmann averaging, 
statistical sampling, 
and the evolution of the thermal density matrix governed by the von Neumann equation. \cite{book_Tannor:2007} 

Recently, an alternative approach to vibrationally resolved electronic spectroscopy, called coherence thermofield dynamics, has been introduced. \cite{Begusic_Vanicek:2020, Begusic_Vanicek:2021, Begusic_Vanicek:2021a, Zhang_Vanicek:2024}
Originally developed in quantum field theory, \cite{Suzuki:1985, Takahashi_Umezawa:1996} thermofield dynamics maps the thermal density operator into a pure state of doubled dimensions, thereby circumventing the more challenging task of propagating the density matrix using the von Neumann equation.
Remarkably, the quantum evolution of the density matrix is exactly equivalent to the propagation of a ``temperature-dependent pure state'' governed by a time-dependent Schr\"odinger equation with an augmented Hamiltonian. 
This ``thermofield dynamics'' has been applied beyond theoretical physics in various fields, such as molecular quantum dynamics \cite{Borrelli_Gelin:2016, Borrelli_Gelin:2017, Gelin_Borrelli:2017, Chen_Zhao:2017, Borrelli_Gelin:2021, Fischer_Saalfrank:2021, Gelin_Borrelli:2023} and electronic structure problems. \cite{Harsha_Scuseria:2019}

Within the Born--Oppenheimer approximation, vibronic spectra are determined exclusively by the coherence components of the density matrix.
In contrast to general thermofield dynamics,\cite{Suzuki:1985, Borrelli_Gelin:2016} the coherence thermofield dynamics \cite{Begusic_Vanicek:2020, Zhang_Vanicek:2024} requires propagation with two different Hamiltonians, namely the vibrational Hamiltonians $H_{g}$ and $H_{e}$ associated with the ground and excited electronic potential energy surfaces.
Yet, like the usual thermofield dynamics, the coherence thermofield dynamics transforms the problem of evaluating a thermal dipole time correlation function into a simpler calculation of a wavepacket autocorrelation function in an augmented configuration space.
This powerful simplification enables simulation of spectra at nonzero temperatures using existing codes for zero-temperature simulations.
The coherence thermofield dynamics idea is completely general and can be combined with any quantum, \cite{Zhang_Vanicek:2024} 
semiclassical,\cite{Begusic_Vanicek:2020, Begusic_Vanicek:2021, Begusic_Vanicek:2021a} 
and other approximate methods. 
An earlier and closely related approximation was used for computing spectrum by propagating a Gaussian density in harmonic potentials. \cite{Reddy_Prasad:2015} 
In weakly anharmonic systems, the combination of coherence thermofield dynamics with the thawed and extended thawed Gaussian approximations
has facilitated the study of
non-Condon effects, \cite{Begusic_Vanicek:2020}
two-dimensional electronic spectra, \cite{Begusic_Vanicek:2021} 
and internal conversion rates. \cite{Wenzel_Mitric:2023a}
However, as in the zero-temperature case, single-trajectory methods face limitations in highly anharmonic systems, thereby constraining their accuracy.

To address this limitation, we introduce the Herman--Kluk coherence thermofield dynamics, a method designed to simulate vibronic spectra at nonzero temperatures while maintaining accuracy in strongly anharmonic systems. 
To this end, we combine general coherence thermofield dynamics with the Herman--Kluk approximation and derive the ordinary differential equations needed for the numerical evaluation.
We demonstrate how the Herman--Kluk coherence thermofield dynamics can be obtained from individual classical trajectories and show that, as a result, its computational cost is comparable to that of evaluating two zero-temperature spectra.
We discuss two natural choices for the width matrices of the frozen Gaussians used in the Herman--Kluk approximation and show that only one choice corresponds to Boltzmann averaging.
To validate our method, we compare the spectra obtained with the Herman--Kluk, thawed-Gaussian, and exact quantum coherence thermofield dynamics in Morse systems at various temperatures.
Moreover, we compare the spectrum obtained from Herman--Kluk coherence thermofield dynamics to the Boltzmann-averaged spectrum computed by propagating excited vibrational states using the Herman--Kluk approximation.

\section{Theory}

Within the Condon approximation, the rotationally averaged absorption cross-section $\sigma(\omega)$ associated with vibronic transitions from the ground electronic state $\vert g \rangle$ to the excited electronic state $\vert e \rangle$ can be computed as the Fourier transform \cite{Heller:1981a, book_Mukamel:1999, book_Tannor:2007, book_Heller:2018}
\begin{align}
    \sigma(\omega) = \frac{2\pi\omega}{3\hbar c} \int_{-\infty}^{\infty} C_{\mu \mu}(t) e^{i\omega t} \, dt
    \label{EQ:Spectra}
\end{align}
of the dipole time correlation function
\begin{align}\label{EQ:Correlation_function}
\begin{split}
    C_{\mu \mu}(t) &= \mathrm{Tr}[\hat{\mu}^{\dagger} \exp( - i\hat{H}_{e}t/\hbar) \hat{\mu}\hat{\rho} \exp( i\hat{H}_{g}t/\hbar) ]
    \\& = \mu^2 C(t),
\end{split}
\end{align}
where 
\begin{align}\label{EQ:Coherence}
    C(t) = \mathrm{Tr}[  \exp( - i\hat{H}_{e}t/\hbar)\hat{\rho} \exp( i\hat{H}_{g}t/\hbar) ]
\end{align}
is the coherence between states $g$ and $e$.
In Eqs.~(\ref{EQ:Correlation_function}) and (\ref{EQ:Coherence}), $\hat{\rho}$ is the vibrational density operator associated with the ground-state Hamiltonian $\hat{H}_{g}$, $\hat{H}_{e}$ denotes the Hamiltonian of the excited state, $\hat{\mu} = \vert\vert  \hat{\vec{\mu}}_{eg} \vert\vert $ (hence $\mu = \vert\vert \vec{\mu}_{eg} \vert \vert)$ is the magnitude of the  coordinate-independent transition dipole moment vector operator $\hat{\vec{\mu}}_{ge}$ between states $g$ and $e$.
Assuming that there are $D$ nuclear degrees of freedom, we can write the ground- and excited-state Hamiltonians as
\begin{align}
    \hat{H}_{j} = \frac{1}{2}\hat{p}^T \cdot m^{-1} \cdot \hat{p} + V_{j}(\hat{q}), \qquad j = e \text{ or } g,
\end{align}
with a $D\times D$ real symmetric mass matrix $m$, the $D$-component momentum operator $\hat{p}$, and scalar potential energy operator $V_{j}(\hat{q})$.

At nonzero temperature $T$, the equilibrium vibrational density operator, 
\begin{align}
    \hat{\rho} = e^{-\beta \hat{H}_{g}} / \mathrm{Tr}(e^{-\beta \hat{H}_{g}}),
\end{align}
follows the Boltzmann distribution
with the scaled inverse temperature $\beta= 1/(k_{B} T)$.
The temperature-dependent spectrum (\ref{EQ:Spectra}) can be obtained from the Boltzmann-averaged time correlation function
\begin{align}\label{EQ:Sum_over_correlation}
    C(t) =  \sum_{j=0}^{\infty} P_{j} e^{i\omega_{g,j}t} \langle \phi_{j} \vert e^{-i\hat{H}_{e} t/\hbar }\vert  \phi_{j} \rangle,
\end{align}
where $\phi_{j}$ is the $j$-th eigenfunction of $\hat{H}_{g}$ with vibrational energy $\hbar \omega_{g,j}$ and occupation probability
\begin{align}
    P_{j} = e^{-\beta \hbar \omega_{g,j}} / \sum_{k=0}^{\infty} e^{-\beta \hbar \omega_{g,k}}.
\end{align}
The evaluation of Eq.~(\ref{EQ:Sum_over_correlation}) for polyatomic molecules becomes computationally expensive because for each contributing vibrational level $j$ (i.~e., where $P_{j}$ is non-negligible), the corresponding high-dimensional wavepacket $\phi_{j}$ must be propagated on the excited-state surface, and the overlaps in Eq.~(\ref{EQ:Sum_over_correlation}) must be computed for each $j$ separately.
When the number of contributing vibrational levels $j$ is very large, one must resort to statistical sampling of Eq.~(\ref{EQ:Sum_over_correlation}). \cite{Manthe_Larranaga:2001, Baiardi_Barone:2013, Wernbacher_Gonzalez:2021}

Instead, in coherence thermofield dynamics, \cite{Begusic_Vanicek:2020, Zhang_Vanicek:2024} the coherence (\ref{EQ:Coherence}) is rewritten \emph{exactly} as the wavepacket autocorrelation function
\begin{align}\label{EQ:Coherence_correlation_function}
    C(t) = \langle \bar{\psi}_{0} \vert e^{-i\hat{\bar{H}}t/\hbar } \vert  \bar{\psi}_{0} \rangle
\end{align}
of the initial thermofield wavepacket $\bar{\psi}_{0}(\bar{q})=\langle q  \vert \hat{\rho}^{1/2}\vert q^{\prime} \rangle$ represented here in the augmented configuration space $\bar{q}=(q, q^{\prime}).$ 
The augmented thermofield Hamiltonian
\begin{align}\label{EQ:TF_Hamiltonian}
    \hat{\bar{H}} =  \hat{\bar{p}}^T \cdot \bar{m}^{-1} \cdot \hat{\bar{p}}/2 + \bar{V}(\hat{\bar{q}})
\end{align}
depends on the augmented momentum $\hat{\bar{p}}=(\hat{p},\hat{p}^{\prime})$ and position $\hat{\bar{q}}=(\hat{q},\hat{q}^{\prime})$ operators, the augmented mass matrix $\bar{m}=\mathrm{diag}(m^{-1}, -m^{-1})$, and the augmented potential $\bar{V}(\hat{\bar{q}})= V_{e}(\hat{q})  -V_{g}(\hat{q}^{\prime}).$
Thus, the Hamiltonian $\bar{H}(\bar{p},\bar{q})=H_{e}(p,q)-H_{g}(p^{\prime}, q^{\prime})$ is separable in the unprimed $p$, $q$ and primed $p^{\prime}$, $q^{\prime}$ degrees of freedom. 
Functions, operators, vectors, and matrices that act on the augmented thermofield space are denoted by bars.

In contrast to Kubo averaging, \cite{book_Kubo:1991, Tuckerman:2010} where the dipole time correlation function is averaged over all possible splits of the Boltzmann distribution $\hat{\rho}$, the derivation of Eq.~(\ref{EQ:Coherence_correlation_function}) uses a symmetric split \cite{Sun_Miller:1998, Wang_Miller:1998, Sun_Miller:1999, Zhao_Miller:2002} of $\hat{\rho}$ to express the dipole time correlation function specifically as a wavepacket autocorrelation function.

\subsection{Herman--Kluk coherence thermofield dynamics}\label{Sec:2}
The remarkably simple formulation of Eq.~(\ref{EQ:Coherence_correlation_function}) makes nonzero-temperature calculations possible with any quantum or semiclassical method designed for zero-temperature calculations.
For example, the efficient thawed Gaussian coherence thermofield dynamics \cite{Begusic_Vanicek:2020} employs the local harmonic approximation for the excited-state potential, while the numerically exact quantum coherence thermofield dynamics \cite{Zhang_Vanicek:2024} uses an expensive split-operator Fourier method.

To capture more anharmonicity than with the thawed Gaussian dynamics without resorting to expensive quantum dynamics, we replace the exact evolution operator in Eq.~(\ref{EQ:Coherence_correlation_function}) with the semiclassical Herman--Kluk propagator \cite{Herman_Kluk:1984} and obtain
\begin{align}\label{EQ:Coherence_HK}
    C(t) = (2\pi\hbar)^{-2D} \int \bar{R}_{t}(\bar{z}_{0}) e^{i\bar{S}_{t}(\bar{z}_{0})/\hbar} \langle \bar{g}_{0} \vert \bar{\psi}_{0} \rangle  \langle \bar{\psi}_{0} \vert \bar{g}_{t} \rangle \,d^{4D}\bar{z}_{0}.
\end{align}
In position representation, the frozen Gaussians $\bar{g}_{t}$ are 
\begin{align}\label{EQ:Gaussian}
    \bar{g}_{t}(\bar{q}) =  \left[\frac{\det{\bar{\Gamma}}}{(\pi\hbar)^{2D}}\right]^{1/4} \exp{\left[\left(- \bar{x}^T \cdot \bar{\Gamma} \cdot \bar{x}/2 + i \bar{p}^T_{t} \cdot \bar{x} \right)/\hbar \right]},
\end{align}
where $\bar{x}:=\bar{q}-\bar{q}_{t}$ is the shifted position vector
and $\bar{\Gamma}$ is a $2D\times 2D$ constant real width matrix.
The centers $\bar{z}_{0} = (\bar{q}_{0},\bar{p}_{0})=(q_{0} , q^{\prime}_{0}, p_{0} , p^{\prime}_{0})$ and $\bar{z}_{t}=(\bar{q}_{t}, \bar{p}_{t})=(q_{t}, q^{\prime}_{t}, p_{t}, p^{\prime}_{t})$ are the initial and final coordinates of a classical trajectory in the augmented, ``double'' phase space, and $\bar{S}_{t}(\bar{z}_{0})$ is the classical action along this trajectory.
The augmented Herman--Kluk prefactor $\bar{R}_{t}(\bar{z}_{0})$ becomes
\begin{align}\label{EQ:HK_prefactor}
\begin{split}
    \bar{R}_{t}(\bar{z}_{0}) = 2^{-D} 
    \det &\left( \bar{M}_{\bar{q}\bar{q}} + \bar{\Gamma}^{-1} \cdot \bar{M}_{\bar{p}\bar{p}}\cdot \bar{\Gamma} \right.
    \\&\left. - i \bar{M}_{\bar{q}\bar{p}} \cdot \bar{\Gamma} + i \bar{\Gamma}^{-1} \cdot \bar{M}_{\bar{p}\bar{q}} \right)^{1/2},
\end{split}
\end{align}
where the $2D\times 2D$ matrices $\bar{M}_{\alpha\beta}$, $\alpha,\beta \in \{\bar{q},\bar{p}\}$, are blocks of the $4D\times 4D$ stability matrix $\bar{M}_{t}(\bar{z}_{0}) = \partial \bar{z}_{t}/ \partial \bar{z}_{0}$.

Phase-space coordinates $(\bar{q}_{t}, \bar{p}_{t})$, classical action $\bar{S}_{t}(\bar{z}_{0})$, and stability matrix $\bar{M}_{t}(\bar{z}_{0})$ along the trajectory satisfy the ordinary differential equations
\begin{align}
    \dot{\bar{q}}_{t} &= \bar{m}^{-1} \cdot \bar{p}_{t}, \label{EQ:ODE1}
    \\  \dot{\bar{p}}_{t} &= -\mathrm{grad\,}\bar{V}(\bar{q}_{t}), \label{EQ:Appendix1}
    \\ \dot{\bar{S}}_{t} (\bar{z}_{0} ) &= \frac{1}{2}\bar{p}_{t}^T \cdot \bar{m}^{-1} \cdot \bar{p}_{t} - \bar{V}(\bar{q}_{t}),\label{EQ:Appendix2}
    \\ \dot{\bar{M}}_{t} (\bar{z}_{0} ) & = \begin{pmatrix}
    0 & \bar{m}^{-1} \\  -\mathrm{Hess\,}\bar{V}(\bar{q}_{t}) & 0
\end{pmatrix} \cdot \bar{M}_{t} (\bar{z}_{0} ), \label{EQ:ODE2}
\end{align}
with initial conditions $(\bar{q}_{t}, \bar{p}_{t})\vert_{t=0} = (\bar{q}_{0}, \bar{p}_{0})$, $\bar{S}_{0}(\bar{z}_{0})=0$, and $\bar{M}_{0}(\bar{z}_{0}) = \mathrm{Id}_{4D}$, where $\mathrm{Id}_{4D}$ is the $4D$-dimensional identity matrix.

Because the augmented Hamiltonian $\bar{H}$ in Eq.~(\ref{EQ:TF_Hamiltonian}) is separable in the spatial dimensions $q $ and $q^{\prime}$, the ordinary differential equations~(\ref{EQ:ODE1})-(\ref{EQ:ODE2}) are also separable.
Therefore, the augmented phase-space coordinates $\bar{z}_{t}=(\bar{q}_{t}, \bar{p}_{t})$, classical action $\bar{S}_{t}(\bar{z}_{0})$ and stability matrix $\bar{M}_{t}(\bar{z}_{0})$ can be reconstructed from the independent phase-space coordinates $z_{t} = (q_{t},p_{t})$, $z^{\prime}_{t} = (q^{\prime}_{t}, p^{\prime}_{t})$, and their corresponding classical actions $S_{t}(z_{0})$, $S^{\prime}_{t}(z_{0}^{\prime})$ and stability matrices $M_{t}(z_{0}) = \partial z_{t} / \partial z_{0}$, $M^{\prime}_{t}(z^{\prime}_{0}) = \partial z_{t}^{\prime} / \partial z_{0}^{\prime}$.
If we group the four parameters $q,p,S,M$ into a single parameter $\Lambda:=(q,p,S,M)$, then the equations for $\Lambda_{t}$ are Eqs.~(\ref{EQ:ODE1})-(\ref{EQ:ODE2}), where $\bar{\Lambda}_{t}$, $\bar{m}$, and $\bar{V}$ are replaced by $\Lambda_{t}$, $m$ and $V_{e}$, with initial conditions $(q_{t}, p_{t})\vert_{t=0} = (q_{0}, p_{0})$, $S_{0}(z_{0})=0$ and $M_{0}(z_{0}) = \mathrm{Id}_{2D}$.
The equations for $\Lambda^{\prime}_{t}$ are Eqs.~(\ref{EQ:ODE1})-(\ref{EQ:ODE2}), where $\bar{\Lambda}_{t}$, $\bar{m}$, and $\bar{V}$ are replaced by $\Lambda^{\prime}_{t}$, $-m$ and $-V_{g}$, with initial conditions $(q^{\prime}_{t}, p^{\prime}_{t})\vert_{t=0} = (q^{\prime}_{0}, p^{\prime}_{0})$, $S^{\prime}_{0}(z^{\prime}_{0})=0$ and $M^{\prime}_{0}(z^{\prime}_{0}) = \mathrm{Id}_{2D}$.
In particular, classical actions along $z_{t}$ and $z_{t}^{\prime}$ follow the equations
\begin{align}
    \dot{S}_{t}(z_{0}) &= p_{t}^T \cdot m^{-1} \cdot p_{t}/2 - V_{e}(q_{t}),
    \\ \dot{S}^{\prime}_{t}(z_{0}^{\prime}) & = - (p_{t}^{\prime})^T \cdot m^{-1} \cdot p_{t}^{\prime}/2 + V_{g}(q^{\prime}_{t}).
\end{align}
Therefore,
$\dot{\bar{S}}_{t}(\bar{z}_{0}) = \dot{S}_{t}(z_{0}) + \dot{S}^{\prime}_{t}(z_{0}^{\prime})$ and $\bar{S}_{t}(\bar{z}_{0}) = S_{t}(z_{0}) + S^{\prime}_{t}(z^{\prime}_{0})$.
The augmented stability matrix is
\begin{align}
    \bar{M}=\begin{pmatrix}
        M_{qq} & 0 & M_{qp}  &0
        \\ 0 & M^{\prime}_{qq} & 0 & M^{\prime}_{qp}
        \\  M_{pq} & 0 & M_{pp} & 0
        \\ 0 & M^{\prime}_{pq} & 0 & M^{\prime}_{pp}
    \end{pmatrix},
\end{align}
where $M_{\alpha\beta}:=\partial \alpha_{t}/ \partial \beta_{0}$ and $M^{\prime}_{\alpha\beta}:=M_{\alpha^{\prime}\beta^{\prime}}$, $\alpha,\beta \in\{ q, p\}$, are the $D\times D$-dimensional blocks of the stability matrices $M$ and $M^{\prime}$.

\subsection{Choice of the width matrix}\label{Sec:Choice_of_width_matrix}

To describe the initial state, which is localized near the minimum of the ground-state potential $V_{g}$, we approximate $V_{g}$ near its minimum by a harmonic approximation $V_{g}(q^{\prime}) = (q^{\prime})^T \cdot  \kappa \cdot q^{\prime}/2$ with a $D\times D$-dimensional force constant matrix $\kappa$.
The initial thermofield wavepacket $\bar{\psi}_{0}(\bar{q}) = \langle q \vert  \hat{\rho}^{1/2} \vert q^{\prime} \rangle = \bar{g}_{i}(\bar{q}) $ is then a Gaussian centered at $\bar{z}_{i}=0$ and characterized by the width matrix
\begin{align}\label{EQ:Widthmatrix}
   \bar{\gamma} := \begin{pmatrix} A & B \\ B & A \end{pmatrix}.
\end{align}
The $D\times D$-dimensional block matrices $A$ and $B$ are temperature-dependent and given by \cite{Begusic_Vanicek:2020} $A = m^{1/2}\cdot \Omega \cdot \mathrm{coth}( \beta \hbar \Omega/2) \cdot m^{1/2}$ and $B = -m^{1/2}\cdot \Omega \cdot \mathrm{sinh}( \beta \hbar \Omega/2)^{-1} \cdot m^{1/2}$, where $\Omega = (m^{-1/2} \cdot \kappa \cdot m^{-1/2})^{1/2}$.

The width matrices $\bar{\Gamma}$ of the Gaussians $\bar{g}_{t}$ in Eq.~(\ref{EQ:Coherence_HK}) can be chosen arbitrarily as in zero-temperature calculations. \cite{Kay:2006, Lasser_Lubich:2020} 
Nevertheless, there are two natural options:

1. $\bar{\Gamma} = \bar{\gamma}$: 
The Gaussians $\bar{g}_{t}$ in Eq.~(\ref{EQ:Coherence_HK}) are temperature-dependent and they couple the excited-state $(q)$ and ground-state $(q^{\prime})$ degrees of freedom, i.~e., $\bar{g}_{t}(\bar{q}) \neq g_{t}(q)g^{\prime}_{t}(q^{\prime})$, where $g_{t}$ and $g_{t}^{\prime}$ are $D$-dimensional Gaussians centered at $z_{t}$ and $z_{t}^{\prime}$, respectively.
Even though the Hamiltonian $\bar{H}$ is separable in $q$ and $q^{\prime}$, the block Toeplitz form~(\ref{EQ:Widthmatrix}) of $\bar{\Gamma}$, with nonzero off-diagonal blocks $B$, prevents a simplification of the Herman--Kluk prefactor~(\ref{EQ:HK_prefactor}) into the product $\bar{R}_{t}(\bar{z}_{0}) = R_{t}(z_{0}) R_{t}^{\prime}(z_{0}^{\prime})$ of two individual ``standard'' Herman--Kluk prefactors \cite{Herman_Kluk:1984, Lasser_Lubich:2020} $R_{t}$ and $R_{t}^{\prime}$, 
because the matrix in the determinant of Eq.~(\ref{EQ:HK_prefactor}) is not block-diagonal.
However, these simplifications into products are necessary to be equivalent to Boltzmann averaging, because the Boltzmann-averaged time correlation function~(\ref{EQ:Sum_over_correlation}) can be written as
\begin{align}
\begin{split}\label{EQ:C_AA}
    C(t) &= (2\pi\hbar)^{-2D} \int R_{t}(z_{0}) R^{\prime}_{t}(z^{\prime}_{0}) e^{i[S_{t}(z_{0}) + S^{\prime}_{t}(z^{\prime}_{0})]/\hbar} 
    \\& \times \langle g_{0} \vert \rho^{1/2} \vert g^{\prime}_{0}\rangle  \langle g^{\prime}_{t} \vert (\rho^{1/2})^{\dagger} \vert g_{t} \rangle \,dz_{0} \,dz^{\prime}_{0},
\end{split}
\end{align}
by transforming Eq.~(\ref{EQ:Sum_over_correlation}) back to Eq.~(\ref{EQ:Coherence}), splitting the density operator symmetrically, and then replacing the exact evolution operators by the Herman--Kluk approximation.

2. $\bar{\Gamma} = \mathrm{diag}(\gamma, \gamma)$, with $\gamma = m^{1/2} \cdot \Omega \cdot m^{1/2}$, even at $T\neq 0$ [This option is equivalent to option $1$ only at zero temperature, when $\bar{\Gamma} = \bar{\gamma} = \mathrm{diag}(\gamma, \gamma)$].
Because $\bar{\Gamma}$ is block-diagonal, $\bar{g}_{t}(\bar{q})= g_{t}(q)g_{t}^{\prime}(q^{\prime})$ with $g_{t}$ and $g_{t}^{\prime}$ centered in $z_{0}$ and $z_{0}^{\prime}$, respectively. 
Furthermore, the block-diagonal structure of the block matrices of $\bar{M}$ permits the factorization of the prefactor as $\bar{R}_{t}(\bar{z}_{0}) = R_{t}(z_{0})R^{\prime}_{t}(z_{0}^{\prime})$.
Due to the separability, equation~(\ref{EQ:Coherence_HK}) can be rewritten in the form of Eq.~(\ref{EQ:C_AA})
and it coincides with the Boltzmann-averaged time correlation function~(\ref{EQ:Sum_over_correlation}), when the exact evolution operator is replaced by the Herman--Kluk propagator using frozen Gaussians with width matrix $\gamma$.
Moreover, Eq.~(\ref{EQ:C_AA}) coincides with those used in Herman--Kluk simulation of infrared spectra at nonzero temperatures (except that there $H_{e} = H_{g} = H$). \cite{Sun_Miller:1998, Wang_Miller:1998, Sun_Miller:1999, Gelabert_Miller:2001, Miller:2001, Wang_Miller:2001a, Sun_Miller:2002, Zhao_Miller:2002, Tao_Miller:2011, Makri:2002, Makri_Miller:2002, Kegerreis_Makri:2007, Church_Ananth:2019, Tatchen_Pollak:2009, Lanzi_Conte:2024}

Because of this separability and explicit connection to Boltzmann averaging, in most of our simulations (except in Fig.~\ref{Fig:3}) we chose the second option, $\bar{\Gamma} = \mathrm{diag}(\gamma, \gamma)$ with $\gamma = m^{1/2} \cdot \Omega \cdot m^{1/2}$.
A further investigation of the choice of the width matrix will be subject to future research.

\subsection{Choice of the sampling density and propagation of trajectories}
The initial conditions $\bar{z}_{0}$ can depend on $T$ through a properly chosen temperature-dependent sampling that accelerates convergence.
When the temperature-dependent density 
\begin{align}\label{EQ:Husimi}
    \bar{\rho}(\bar{z}_{0})= (2\pi\hbar)^{-2D} \vert \langle \bar{\psi}_{0} \vert \bar{g}_{0} \rangle \vert^{2}
\end{align}
is employed for Monte Carlo integration of Eq.~(\ref{EQ:Coherence_HK}), its estimator is exact at the initial time $t=0$ for any number of trajectories (because the estimation is $1$ at $t=0$).

Temperature-dependent sampling is best if one is interested in spectra at only a few temperatures.
In our simulations, we therefore sample initial conditions $\bar{z}_{0}$ from Eq.~(\ref{EQ:Husimi}) for each temperature $T$ individually.

We note, however, that for multiple simulations spanning a narrow temperature range, it may be more efficient to reuse the same ensemble of trajectories, with initial conditions sampled from a density based on the average temperature within the range.
If the temperature range is broad and one would still want to employ a single set of trajectories for all simulations, a density based on a higher temperature should be employed to ensure an adequate coverage of rare events.

The propagation of trajectories is independent of temperature. 
Among various numerical methods for solving the ordinary differential equations~(\ref{EQ:ODE1})-(\ref{EQ:ODE2}), we chose the second-order St\o rmer-Verlet algorithm \cite{Brewer_Hulme:1997, Hairer_Wanner:2003, book_Hairer_Wanner:2006, Lasser_Sattlegger:2017} to preserve the symplectic structure of the underlying classical Hamiltonian system.
In single-trajectory coherence thermofield dynamics methods, \cite{Begusic_Vanicek:2020} $q_{0}^{\prime}$ is located at the minimum of $V_{g}$ with $p_{0}^{\prime}=0$.
Consequently, $(q_{t}^{\prime}, p_{t}^{\prime})=(q_{0}^{\prime}, p_{0}^{\prime}) \equiv \mathrm{constant}$, and neither position nor momentum need to be propagated.
In contrast, in the multi-trajectory Herman--Kluk coherence thermofield dynamics, the initial conditions $\bar{z}_{0}=(z_{0},z^{\prime}_{0})$ are sampled. \cite{Kroeninger_Lasser_Vanicek:2023, Kroeninger_Lasser_Vanicek:2024} 
With probability one, $q_{0}^{\prime}$ is not exactly at the minimum of $V_{g}$ and momentum $p_{0}^{\prime}$ is not zero. 
Therefore, not only $z_{t}$ but also $z_{t}^{\prime}$ (along with their classical actions and stability matrices) must be propagated numerically on $V_{e}$ but also on $-V_{g}$, respectively.
Because of the independent propagation of ``physical'' $(z_{t}, S_{t}, M_{t})$ and ``fictitious'' $(z_{t}^{\prime}, S_{t}^{\prime}, M_{t}^{\prime})$ trajectories, parameters $\bar{z}_{t}$, $\bar{S}_{t}$, and $\bar{M}_{t}$ of trajectories in doubled phase space can be easily reconstructed from the single phase-space trajectories. 
Hence, the computational cost of the Herman--Kluk coherence thermofield dynamics is approximately equal to that of two zero-temperature Herman--Kluk simulations.

\section{Numerical examples}
In this section, we compare absorption spectra computed from Herman--Kluk coherence thermofield dynamics with those obtained from either the exact quantum \cite{Zhang_Vanicek:2024} or thawed-Gaussian \cite{Begusic_Vanicek:2020} coherence thermofield dynamics.

\begin{figure*}
   \centering
    \includegraphics{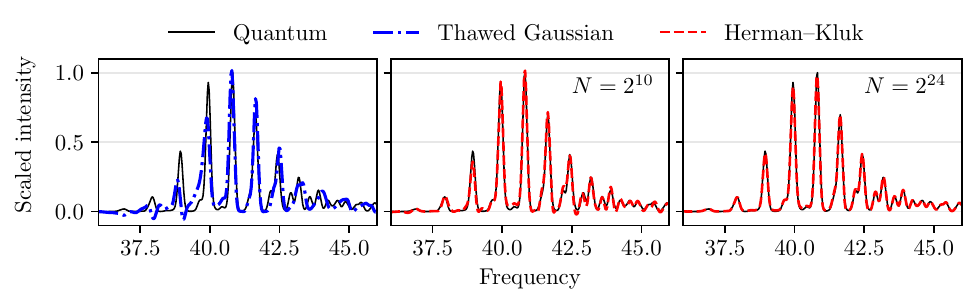}
    \caption{Spectra at scaled temperature $T_{\omega} = 1$ obtained from Herman--Kluk, thawed Gaussian, and quantum coherence thermofield dynamics in a Morse system~(\ref{EQ:Morse}) with anharmonicity parameter $\chi=0.02$ and reference position $q_{\rm ref}=1.5$.
    The Herman--Kluk results were obtained using either $2^{10}=1024$ (middle panel) or $2^{24} \approx 16.8\cdot 10^{6} $ trajectories (right panel).
   All spectra were scaled by the maximum intensity of the quantum coherence thermofield dynamics. }
    \label{Fig:1}
\end{figure*}

Our one-dimensional example consists of a ground-state harmonic potential $V_{g}(q^{\prime}) = (q^{\prime})^T\cdot \kappa \cdot q^{\prime}/2$, $\kappa =1$, and an excited-state Morse potential,
\begin{align}\label{EQ:Morse}
    V_{e}(q) = V_{0} + \frac{\omega_{e}}{4\chi}\left[ 1- e^{-\sqrt{2m\omega_{e} \chi} (q-q_{\rm ref})} \right]^{2},
\end{align}
with equilibrium position $q_{\rm ref}$, minimum energy $V_{0}  = 40$, harmonic frequency $\omega_{e} = 0.9$,  and varying dimensionless anharmonicity $\chi$.
For the propagation, we use a time step of $0.1$ for a total of $2000$ steps. We use natural units, where $m=\hbar = k_{B} = \mu =1$, and apply a Gaussian damping function (with a half-width at half-maximum of 15) to the autocorrelation function $C(t)$ prior to evaluating the spectra via the Fourier transform~(\ref{EQ:Spectra}).

\begin{figure*}
    \centering
    \includegraphics{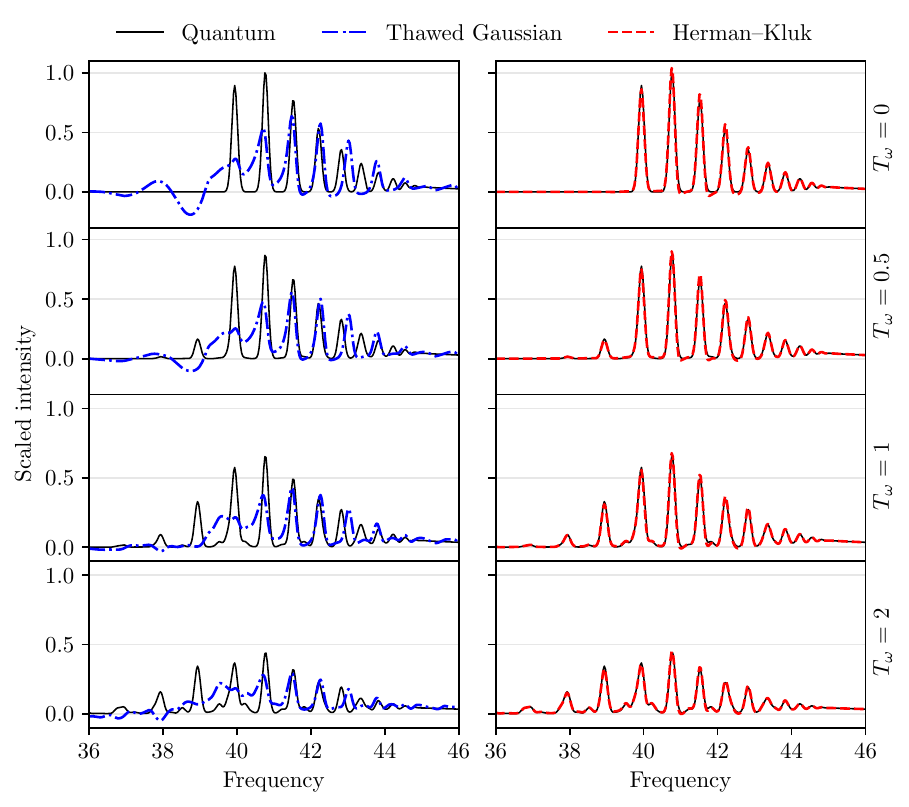}
    \caption{
    Spectra at different scaled temperatures $T_{\omega}$ obtained from Herman--Kluk (with $2^{24}$ trajectories), thawed Gaussian, and quantum coherence thermofield dynamics.
    To increase the anharmonicity, the parameters of the Morse potential from Fig.~\ref{Fig:1} were changed to $\chi=0.04$ and $q_{\rm ref}=1.6$.
    All spectra were scaled such that the maximum intensity of the quantum coherence thermofield dynamics result at $T_{\omega}=0$ is $1$. }
    \label{Fig:2}
\end{figure*}

Figure~\ref{Fig:1} displays the spectra of a Morse system (with $q_{\rm ref} = 1.5$ and anharmonicity $\chi = 0.02$) at the scaled temperature $T_{\omega} = k_{B}T/(\hbar \omega_{g}) = 1$, where $\omega_{g}=\sqrt{\kappa/m}$ is the ground-state vibrational frequency. 
The spectrum obtained from the thawed Gaussian coherence thermofield dynamics deviates from the quantum result but still captures the overall spectral shape, including peak position and intensities, at least qualitatively.
However, there are unphysical negative peaks because the thawed Gaussian approximation can be interpreted as an exact solution to a nonlinear Schr\"{o}dinger equation, i.e., a Schr\"{o}dinger equation with a state-dependent potential, for which positivity of spectrum is not guaranteed. \cite{Lasser_Lubich:2020, Vanicek:2023}
Figure~\ref{Fig:1} shows two spectra obtained with the Herman--Kluk coherence thermofield dynamics using either $2^{10}$ or $2^{24}$ trajectories. 
Despite the first result not being fully converged, it already exhibits excellent agreement with the quantum result, capturing all peaks and their intensities, including asymmetric low-intensity peaks at both low and high frequencies.
Moreover, the negative-peak issue almost disappears.

Figure~\ref{Fig:2} compares spectra at different scaled temperatures $T_{\omega}\in\{0,0.5,1,2\}$ obtained from the dynamics in a Morse potential with larger displacement $(q_{\rm ref}=1.6)$ and increased anharmonicity $(\chi=0.04)$, leading to a lower dissociation energy compared to the potential analyzed in Fig.~\ref{Fig:1}. 

At zero temperature $T_{\omega}=0$, the thawed Gaussian coherence thermofield dynamics already fails, as the spectrum reproduces only a few peaks correctly.
The positions of the first two high-intensity peaks from the quantum result are correctly reproduced but with wrong intensities which might be caused from the overlap with the unphysical large negative peak arising from nonlinearity.
At higher frequencies, the spectrum obtained from the thawed Gaussian coherence thermofield dynamics shows peaks with reasonable intensities but incorrect position.
At $T_{\omega} \in \{0.5,1,2\}$, some peaks are still qualitatively reproduced; however,
new low-intensity peaks (``hot bands'') that emerge at higher temperatures are not captured.
It is possible that these peaks are not captured, because they are cancelled by the large negative peak appearing clearly in the zero-temperature simulation.
In contrast, the Herman--Kluk spectra reproduce quantum spectra almost perfectly at all scaled temperatures $T_{\omega}$. 
In particular, the Herman--Kluk results reproduce the correct peak positions, intensities, shapes and asymmetries even for hot bands (the low-intensity peaks emerging at higher temperatures). 

\begin{figure}
    \centering
    \includegraphics{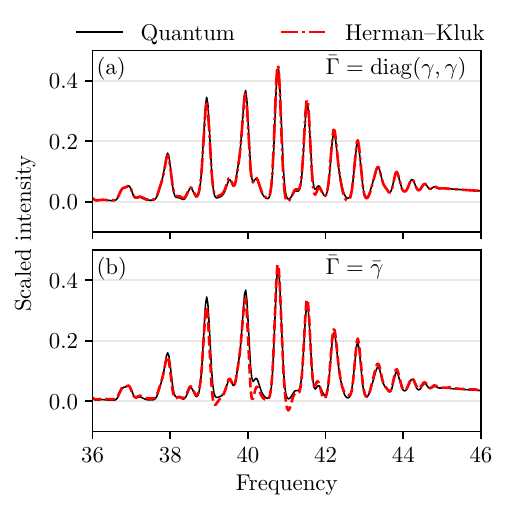}
    \caption{Spectra obtained from Herman--Kluk and quantum coherence thermofield dynamics of the same system as Fig.~\ref{Fig:2} at $T_{\omega} = 2$. The width matrix $\bar{\Gamma}$ of the Gaussians $\bar{g}_{t}$ for the Herman--Kluk coherence thermofield dynamics simulations was either (a) $\bar{\Gamma} = \mathrm{diag}(\gamma, \gamma)$ (upper panel) or (b) $\bar{\Gamma} = \bar{\gamma}$ (lower panel).
    }
    \label{Fig:3}
\end{figure}

Figure~\ref{Fig:3} compares spectra obtained from Herman--Kluk coherence thermofield dynamics using two different choices of the width matrix $\bar{\Gamma}$ of the Gaussians $\bar{g}_{j}$. 
The upper panel, identical to the lowest right-hand panel of Fig.~\ref{Fig:2}, uses frozen Gaussians with zero-temperature width matrix $\bar{\Gamma}=\mathrm{diag}(\gamma, \gamma)$, where $\gamma = m^{1/2} \cdot \Omega \cdot m^{1/2}$ (option $2$ in Sec.~\ref{Sec:Choice_of_width_matrix}). 
The lower panel shows the result for $\bar{\Gamma} = \bar{\gamma}$, i.~e., frozen Gaussians with the same width matrix as the initial state (option $1$ in Sec.~\ref{Sec:Choice_of_width_matrix}).
The result in the lower panel captures the general shape, position, and asymmetry of most peaks, making it an improvement over the thawed Gaussian coherence thermofield dynamics.
However, it is clearly worse than the simulation from the upper panel, as it shows significant deviations from the quantum result near frequencies $39$, $40$, and $41$ $\mathrm{n.u.}$
Additionally, visible negative peaks appear, which can occur in spectra obtained from the Herman--Kluk approximation for the same reason \cite{Lasser_Lubich:2020} as they did in Figs.~\ref{Fig:1} and \ref{Fig:2} for the thawed Gaussian approximation (the Herman--Kluk wavefunction, too, solves exactly a state-dependent nonlinear Schr\"{o}dinger equation).
\begin{figure}
   \centering
    \includegraphics{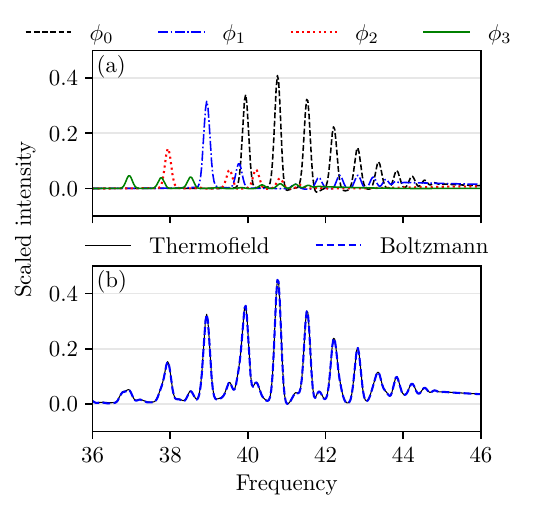}
    \caption{(a) Absorption spectra from initial vibrational levels (up to $j=3$) of the ground electronic state obtained from the Herman--Kluk approximation (each spectrum is already weighted by the Boltzmann factor). Each line was produced with $2^{24}$ trajectories.
    (b) Comparison of the spectrum obtained from thermofield dynamics to the Boltzmann-averaged spectrum from initial vibrational levels (up to $j=6$) of the same system as in Fig.~\ref{Fig:2} at scaled temperature $T_{\omega}=2$.}
    \label{Fig:4}
\end{figure}
Moreover, the Herman--Kluk coherence thermofield dynamics becomes equivalent to Boltzmann averaging~(\ref{EQ:Sum_over_correlation}) when $\bar{\Gamma} = \mathrm{diag}(\gamma, \gamma)$ and when the exact evolution operator of Eq.~(\ref{EQ:Sum_over_correlation}) is replaced by the Herman-Kluk propagator with frozen Gaussians that have width matrix $\gamma$.
In Fig.~\ref{Fig:4}, we therefore compare the Herman--Kluk coherence thermofield spectrum with that obtained by Herman--Kuk Boltzmann averaging, and show the decomposition of the Boltzmann-averaged spectrum into the contributions from different initial vibrational levels.
For each initial vibrational level $j$, the initial conditions were sampled from $\vert\langle \phi_{0} \vert g_z \rangle\vert^{2}$.
We observe very close agreement between the thermofield and Boltzmann-averaged results.
Yet, the thermofield approach accounts for all thermal contributions in a single calculation, whereas Boltzmann averaging requires the individual propagation of each initial vibrational wavepacket.

\section{Conclusion}
We presented the Herman--Kluk coherence thermofield dynamics, a method for obtaining vibronic spectra at nonzero temperatures.
This semiclassical approach offers a middle ground between the efficient but crude thawed-Gaussian \cite{Begusic_Vanicek:2020} and the exact but inefficient quantum \cite{Zhang_Vanicek:2024} coherence thermofield dynamics.

The Herman--Kluk coherence thermofield dynamics is a multi-trajectory method whose trajectories exist in a phase space of doubled dimension. 
Each trajectory is composed of two independent trajectories in single phase space, one evolving on the ground-state surface and the other on the excited-state surface. 
We showed how spectra are obtained from the Herman--Kluk coherence thermofield dynamics using these individual trajectories, their classical actions and their stability matrices. 
Consequently, the propagation cost of the Herman--Kluk coherence thermofield dynamics is the same as the cost of evaluating two zero-temperature spectra within the Herman--Kluk approximation (under the assumption that in each case the same number of trajectories is used).

In our numerical examples, we studied a one-dimensional Morse potential with varying anharmonicities and temperatures.
In the weakly anharmonic case, the spectrum obtained from the thawed Gaussian coherence thermofield dynamics yielded a qualitatively reasonable result, while the spectra obtained from the Herman--Kluk coherence thermofield dynamics exhibited almost no deviation from the exact quantum result.
In the more anharmonic case and at increasing temperatures, the spectra from thawed Gaussian coherence thermofield dynamics failed to reproduce emerging hot bands, whereas the Herman--Kluk coherence thermofield dynamics maintained high accuracy. 
To explore the origin of the extended spectral range at high temperatures, we compared the Herman--Kluk coherence thermofield dynamics with its Boltzmann-averaged counterpart.
While Boltzmann-averaging requires computing multiple wavepacket autocorrelation functions, each corresponding to a different initial vibrational level, the spectrum obtained from the Herman--Kluk coherence thermofield dynamics requires the evaluation of a single wavepacket autocorrelation function at a cost similar to evaluating two zero-temperature wavepacket autocorrelation functions.

The coherence thermofield dynamics concept is general and can be combined with any dynamical method. 
In the Herman--Kluk case, the freedom of choice of the width matrix of the frozen Gaussians results in several inequivalent methods. 
We showed that only one agrees with the Boltzmann averaging.
It turns out that this method also agrees with double phase-space approaches that have been previously used for evaluating infrared spectra and rate constants.\cite{Sun_Miller:1998, Wang_Miller:1998, Sun_Miller:1999, Gelabert_Miller:2001, Miller:2001, Wang_Miller:2001a, Sun_Miller:2002, Zhao_Miller:2002, Tao_Miller:2011, Makri:2002, Makri_Miller:2002, Kegerreis_Makri:2007, Church_Ananth:2019, Tatchen_Pollak:2009, Lanzi_Conte:2024}

The Herman--Kluk thermofield wavepacket autocorrelation function presented here retains the ``standard'' form of a zero-temperature Herman--Kluk wavepacket autocorrelation function. 
It includes one Herman--Kluk prefactor, one classical action and various overlaps of the initial state with Gaussians. 
Hence, it may easily be combined with further approximations such as Filinov filtering, \cite{Filinov:1986, Makri_Miller:1988, Walton_Manolopoulos:1996} hybrid and cellular dynamics, \cite{Grossmann:2006,Goletz_Grossmann:2009,Grossmann:2016} time-averaging, \cite{Elran_Kay:1999a,Kaledin_Miller:2003,Buchholz_Ceotto:2016,Buchholz_Ceotto:2018} or multiple coherent states, \cite{Ceotto_Aspuru-Guzik:2009a} 
to further enhance its computational efficiency. 

\section*{Acknowledgments}
F. Kr\"oninger and J. Van\'{i}\v{c}ek acknowledge the financial support from the EPFL.

\bibliographystyle{aipnum4-2}
\bibliography{Coherence_Thermofield_Herman_Kluk}
\end{document}